\documentclass{article}
\usepackage{hyperref}
\usepackage{graphicx}
\usepackage{amssymb,latexsym,amsmath,amsfonts,amsthm}
\usepackage{xcolor}
\usepackage{bm}
\usepackage[a4paper,left=2cm,right=2cm,top=3cm,bottom=3cm]{geometry}

\newtheorem{theorem}{Theorem}

\title{Constructing the Hamiltonian from the behaviour of a dynamical system by proper symplectic decomposition}

\newcommand{\orcidID}[1]{{\footnotesize{(#1)}}}
\newcommand{\inst}[1]{$^{#1}$}

\author{Nima Shirafkan\inst{1}\orcidID{0000-0002-4011-2029},
	Pierre Gosselet\inst{2}\orcidID{0000-0002-2265-2427},\\
	Franz Bamer\inst{1}\orcidID{0000-0002-8587-6591},
	Abdelbacet Oueslati\inst{2}\orcidID{0000-0003-1403-4781},\\
	Bernd Markert\inst{1}\orcidID{0000-0001-7893-6229} and
	G\'ery de Saxc\'e\inst{2}\orcidID{0000-0002-0961-0513} \\
$(1)$ Institute of General Mechanics, RWTH, Aachen, Germany \\
\url{https://www.iam.rwth-aachen.de} \\
$(2)$ Univ. Lille, CNRS, Centrale Lille, UMR 9013 -- LaMcube --\\ F-59000, Lille, France  \\
\url{http://lamcube.univ-lille.fr}
}

\begin{document}
\maketitle              
\begin{abstract}
The modal analysis is revisited through the symplectic formalism, what leads to two intertwined eigenproblems. Studying the properties of the solutions, we prove that they form a canonical basis. The method is general and works even if the Hamiltonian is not the sum of the potential and kinetic energies. On this ground, we want to address the following problem: data being given in the form of one or more structural evolutions,  we want to construct an approximation of the Hamiltonian from a covariant snapshot matrix and to perform a symplectic decomposition. We prove the convergence properties of the method when the time discretization is refined. If the data cloud is not enough rich, we extract the principal component of the Hamiltonian corresponding to the leading modes, allowing to perform a model order reduction for very high dimension models. The method is illustrated by a numerical example.

{\bf Keywords}: {Symplectic mechanics; Modal analysis; Model order reduction; Principal component analysis.}

Corresponding author email: \href{mailto:gery.de-saxce@univ-lille.fr}{gery.de-saxce@univ-lille.fr}

\end{abstract}

\section{Introduction}
In structural mechanics, the modal analysis coupled with the finite element method is widely used by engineers to determine the eigenmodes and eigenfrequencies of linear dynamical systems \cite{Geradin 2015}.

Very large numerical models are ubiquitous in structural dynamics. Working in high dimension spaces is time-consuming, often intractable and requires storing big pieces of data, hence the need to simplify the models to make them easier and faster to interpret by users. The Proper Orthogonal Decomposition (POD) is one of the most successfully used model reduction technique for nonlinear systems \cite{Chinesta 2014}. Nevertheless, it is based on the metric structure of the configuration space, while for dynamical systems the phase space is naturally equipped with a symplectic structure \cite{Libermann_1987,Souriau_1997b}.

The aim of the present work is, for large scale conservative systems, to develop a new method of Proper Symplectic Decomposition (PSD) able to extract the leading eigenmodes of the modal analysis and the principal component of the Hamiltonian of the system from a data cloud comprised of one or many evolutions of the system subjected to external excitations. 

In modern literature, a PSD-based model reduction technique has been proposed by Peng and. Mohseni in \cite{Peng 2016} where the symplectic projection is determined from a snapshot matrix solving a nonlinear optimization problem for linear systems. Our strategy is to develop an alternative PSD method leading to a linear eigenproblem for linear structures, the nonlinear problem being set aside for the modeling of dissipative systems.

\section{The modal analysis as an equivalence problem}

The phase space $ V = \mathbb{R}^{2 N}$  being endowed with the symplectic form:
\[ \omega (\bm{x}, \bm{x}') = \bm{x}^T \bm{J} \, \bm{x}' = \bm{q}^T \bm{p'} - \bm{p}^T \bm{q}'
\]
where $ \bm{q}$  are the degrees of freedom, $\bm{p}$  are the moments, $\bm{x}^T = (\bm{q}^T, \bm{p}^T)$ and $\bm{J}$ is a skew-symmetric matrix.
The motion of the system is governed by the canonical equations  
\[ \dot{\bm{x}} = \nabla^\omega h = \bm{J} \nabla h
\]
where $\nabla^\omega h $ is the symplectic gradient of the Hamiltonian  $h$ (or Hamiltonian vector field). 
The symplectic matrices  $\bm{S}$ that leave invariant  $\omega$ (\textit{i.e.}  $\bm{S}^T \bm{J} \, \bm{S} = \bm{J}$) form the symplectic group $Sp(2\, N, \mathbb{R})$. 
The dual space $V^*$ is equipped with a Poisson bracket $ \left\lbrace \cdot, \cdot \right\rbrace$ such that $ \left\lbrace f, g \right\rbrace= \omega (\nabla^\omega g, \nabla^\omega f)$.
The modal analysis can be rewritten saying there is a symplectic matrix $\bm{S}$  mapping $h (\bm{x})$ and its Hessian matrix $\bm{H} \in V^* \otimes V^* $ onto the Hamiltonian $h' (\bm{x}')$ of a reduced system of independent harmonic oscillators and its diagonal Hessian matrix $\bm{H}'$. The equivalence problem consists in finding a symplectic matrix $\bm{S}$ such that:
\[ h' = h \circ \bm{S}, \quad \mbox{then} \quad 
    \bm{S}^T \bm{H} \, \bm{S} = \bm{H}'
\]
\section{Intertwined eigenproblems and spectral decomposition}

Introducing the Hamiltonian matrix $\bm{H}_\omega =  \bm{J}^{-1} \bm{H}   \in V \otimes V^* $  and decomposing the symplectic matrix into columns ($\bm{S} = \left[ \bm{u}_1, \cdots, \bm{u}_N, \bm{v}_1, \cdots, \bm{v}_N \right]  $) leads to two \textbf{intertwined eigenproblems}:
\begin{equation}
   \bm{H}_\omega  \bm{u}_i = k_i \bm{v}_i, \quad
     \bm{H}_\omega  \bm{v}_i = - g_i \bm{u}_i
\label{H_omega u = - k v & H_omega v = g u}
\end{equation} 
that can be transformed into a classical eigenproblem:
\begin{equation}
\bm{H}^2_\omega  \bm{u}_i = \lambda_i \bm{u}_i
\label{H^2_omega u_i = lambda_i u_i} 
\end{equation} 
where $\lambda_i = - k_i g_i $. 
The properties of the eigenmodes are given by the following result:
\begin{theorem}
If the Hessian matrix $\bm{H}$ is positive definite:
\begin{itemize}
 \item the eigenvalues $\lambda$ of $\bm{H}^2_\omega $ are negative.
 \item the twin eigenvectors $\bm{u}_i$ and $\bm{v}_i$ of the eigenproblem (\ref{H_omega u = - k v & H_omega v = g u}) are not orthogonal: $\omega (\bm{u}_i, \bm{v}_i) \neq 0$.
 \end{itemize} 
\end{theorem}
Scaling the eigenvectors by 
\begin{equation}
   \omega (\bm{u}_i, \bm{v}_i) = 1
 \label{normalization condition} 
 \end{equation} 
 they form a canonical basis of $V$.

\textbf{Remark 1:} in practice, if the structure has suffisant supports to avert rigid displacements, the Hessian matrix is positive definite. 

\textbf{Remark 2:} for the particular case of the standard modal analysis where the Hamiltonian is decoupled:
\begin{equation}
      h(\bm{x}) 
       = \frac{1}{2} \bm{p}^T \bm{M}^{-1} \bm{p} + \frac{1}{2} \bm{q}^T \bm{K} \bm{q},\quad 
\bm{u}^T_i = \left[ \bm{a}^T_i , \bm{0}^T \right], \quad
\bm{v}^T_i = \left[ \bm{0}^T, (\bm{M}  \bm{a}_i)^T \right] 
\label{u_j = (q_j 0) & v_j (0 M q_j)} 
\end{equation} 
we recover the eigenproblem $ \bm{K} \, \bm{a}_i = \omega^2_i  \bm{M} \, \bm{a}_i$  with $\lambda_i = - \omega^2 _i $ and the normalization condition $\omega (\bm{u}_i, \bm{v}_i) = \bm{a}^T_i \bm{M} \, \bm{a}_i = 1 $.

\textbf{Remark 3:} Our method is more general and allows to treat also cases where there are terms coupling $\bm{q}$ and $\bm{p}$, for instance in problems with 
Coriolis' force or electromagnetic fields.

\vspace{0.3cm}

Another result of interest is:
\begin{theorem}
If $(\bm{u}_1, \cdots, \bm{u}_N, \bm{v}_1, \cdots, \bm{v}_N) $ is a canonical basis, let 
$\bm{u}^*_i = - \bm{J} \, \bm{u}_i$ and $\bm{v}^*_i =  \bm{J} \, \bm{v}_i$, \\then $(\bm{v}^*_1, \cdots, \bm{v}^*_N, \bm{u}^*_1, \cdots, \bm{u}^*_N) $ is its dual basis.
\label{thm dual basis} 
\end{theorem}
Indeed, it leads to the \textbf{spectral decomposition} of the identity of $ \mathbb{R}^{2 N}$, next of the Hamiltonian matrix $\bm{H}_\omega$ and of the Hessian matrix $\bm{H}$:
\[   \bm{I}_{2 N} =  \bm{v}_j \otimes \bm{u}^*_j  + \bm{u}_j \otimes \bm{v}^*_j , \qquad 
\bm{H} _\omega   = g_j \bm{v}_j \otimes \bm{u}^*_j - k_j \bm{u}_j \otimes \bm{v}^*_j 
\]
\begin{equation}
     \bm{H}  = g_j \bm{u}^*_j \otimes \bm{u}^*_j  + k_j \bm{v}^*_j \otimes \bm{v}^*_j 
\label{spectral decomposition of H} 
\end{equation}

\section{Proper symplectic Decomposition (PSD)}

We hope to address the following problem:  data being given in the form of a structural evolution  $\left[ 0, T\right] \rightarrow V : t \mapsto \bm{x} (t)$ (or a concatenation of structural evolutions), we want to construct an approximation of the Hamiltonian of the system. The functional space of the components of these evolutions is endowed with the metrics:
\begin{equation}
    (f_1 \mid f_2) =  \frac{1}{T}  \, \int^T_0 f_1 (t) \, f_2 (t) \, dt
 \label{( f_1 mid f_2 ) = } 
 \end{equation}
Our method consists in decomposing  the interval   $\left[ 0, T \right] $    into $m$  subintervals $I_k$  of timestep $\Delta t_k$ and reference point $t_k \in I_k $. Starting from a structural evolution  $\left[ 0, T\right] \rightarrow V : t \mapsto \bm{x} (t)$ as data, we construct, through the isomorphism $\bm{J}$ from $V$  into its dual $V^*$, the \textbf{covariant snapshot matrix}:
\begin{equation}
 \bm{X}^* = \left[ \bm{J} \bm{x} (t_1), \cdots, \bm{J} \bm{x} (t_m)\right]
 \label{covariant snapshot matrix} 
\end{equation}
representing a map of codomain the dual space  $V^*$ and of domain the Euclidean approximation space $W$ of dimension $m$, equipped with the scalar product  between snapshot vectors $\bm{f}_j = [f_j(t_1), \cdots, f_j (t_m)]^T $ :
\[ (\bm{f}_1  , \bm{f}_2)  = \frac{1}{T} \sum^m_{k = 1} f_1 (t_k) \, f_2 (t_k) \Delta t_k 
\]
a discretized version of (\ref{( f_1 mid f_2 ) = }). The corresponding Gram's matrix of the metrics being:
\[ \bm{G}_t = \frac{1}{T} \,diag (\Delta t_1, \cdots , \Delta t_m)
\]
The symmetric matrix:
\begin{equation}
  \bm{H} = 2 \, \bm{X}^*\, \bm{G}_t \,( \bm{X}^*)^T 
\label{B = C G C^T} 
\end{equation}
is positive semi-definite because the metrics of $W$ is positive:
\[ \bm{x}^T \, \bm{X}^* \bm{G}_t \, (\bm{X}^*)^T \bm{x} 
=  ( (\bm{X}^*)^T \bm{x}  , (\bm{X}^*)^T \bm{x}  )  \geq 0
\]
Next an approximation of the eigenmodes $\bm{u}_i $ and $\bm{v}_i $  of the system can be obtained by solving the eigenproblem (\ref{H^2_omega u_i = lambda_i u_i}).

\section{Convergence properties of the method}

We would like to show that $\bm{H}$ converges to a matrix which allows to find the exact eigenvectors when the time interval of the data increases and the time discretization is refined.  To set these ideas down on a simple problem of standard modal analysis, we consider the  free vibrations of a discrete system with non null initial velocity. According to the modal decomposition, we have:
\[ \begin{aligned}
\bm{q} (t)& = \sum^N_{i = 1} \frac{\dot{q}'_i (0)}{\omega_i}\, \sin (\omega_i t) \, \bm{a}_i
= \sum^N_{i = 1} \alpha_i \, f_i (t) \, \bm{a}_i  , \\ 
\bm{p} (t) &= \bm{M} \, \dot{\bm{q}} (t) 
= \sum^N_{i = 1} \alpha_i \, \dot{f}_i (t) \, \bm{M} \, \bm{a}_i
\end{aligned}
\]
with $f_i (t) = \sin (\omega_i t) $ and $\dot{f}_i (t) = \omega_i \cos (\omega_i t)$. 
The time evolution of the structure in terms of covariant vector is :
\[\bm{x}^* (t) = \bm{J} \, \bm{x} (t) 
  = \left[ {{\begin{array}{*{20}c}
		 \bm{p} (t) \hfill \\
		- \bm{q} (t)\hfill \\
		\end{array} }} \right]
		= \sum^N_{i = 1} \alpha_i \, \left[ \dot{f}_i (t) \,  \bm{v}^*_i - f_i (t) \,  \bm{u}^*_i\right] 
\]
where $(\bm{u}^*_i)^T = \left[ \bm{0}^T , \bm{a}^T_i  \right], \;
(\bm{v}^*_i)^T = \left[(\bm{M}  \bm{a}_i)^T ,  \bm{0}^T   \right]  $
are the elements of the dual basis. Owing to (\ref{covariant snapshot matrix}) and (\ref{B = C G C^T}) and refining the time discretization, one has:
\[  \begin{aligned} \underset{m \to \infty}{lim}\,   \bm{H} &=  \underset{m \to \infty}{lim}\,  \frac{2}{T} \sum^m_{k = 1} \Delta t_k  \bm{x}^* (t_k) \otimes \bm{x}^* (t_k)
        = \frac{2}{T} \int^T_0 \bm{x}^* (t) \otimes \bm{x}^* (t) \, dt
\\&\begin{aligned} = 2 \, \sum^N_{i , j= 1} \alpha_i \, \alpha_j  \, 
&\left[
      (\dot{f}_i \mid \dot{f}_j)  \, \bm{v}^*_i \otimes \bm{v}^*_j 
       - (\dot{f}_i  \mid  f_j ) \, \bm{v}^*_i \otimes \bm{u}^*_j \right.\\&\left.
        -  ( f_i \mid  \dot{f}_j )  \, \bm{u}^*_i \otimes \bm{v}^*_j
        +  (f_i \mid   f_j )  \, \bm{u}^*_i \otimes \bm{u}^*_j 
   \right] \end{aligned}
\end{aligned}\]
where:
\[ (f_i \mid f_i) = \frac{1}{2}\, \left[ 1 - \frac{\sin (2\, \omega_i T)}{2\, \omega_i T}\right], \qquad
 (\dot{f}_i \mid \dot{f}_i) = \frac{\omega^2_i}{2}\, \left[ 1 + \frac{\sin (2\, \omega_i T)}{2\, \omega_i T}\right]
\]
When $T$ approaches $+\infty$, one has:
\[ \underset{T \to \infty}{lim}\,  (f_i \mid f_i) = \frac{1}{2}, \qquad
\underset{T \to \infty}{lim}\,  (\dot{f}_i \mid \dot{f}_i) = \frac{\omega^2_i}{2}
\]
The other scalar product above approaching zero, then it remains:
\[  \underset{T \to \infty}{lim}\,  \underset{m \to \infty}{lim}\,   \bm{H} 
= 2 \, \sum^N_{i= 1} \alpha^2_i  \, [
      (f_i \mid  f_i )  \, \bm{u}^*_i \otimes \bm{u}^*_i
      +  (\dot{f}_i \mid \dot{f}_i ) \, \bm{v}^*_i \otimes \bm{v}^*_i ]
\]
\[  \underset{T \to \infty}{lim}\,  \underset{m \to \infty}{lim}\,   \bm{H} 
= \sum^N_{i= 1} \alpha^2_i  \, [
        \bm{u}^*_i \otimes \bm{u}^*_i  +  \omega^2_i  \bm{v}^*_i \otimes \bm{v}^*_i ]
\]
Comparing to the spectral decomposition (\ref{spectral decomposition of H}) we obtain by identification:
\begin{equation}
 g_i =  \alpha^2_i , \qquad 
      k_i =  \alpha^2_i  \omega^2_i, \qquad 
      \lambda_i = - k_i g_i = - \alpha^4_i  \omega^2_i
\label{g_i = alpha^2_i & k_i = & lambda_i =} 
\end{equation} 

Solving the eigenproblem (\ref{H^2_omega u_i = lambda_i u_i}), we would find the exact orthogonal modes. With sufficiently large values of $T$ and $m$, good approximations of these modes are expected. 

\section{Numerical application}

To illustrate the method, let us consider an elastic truss of length $L$, cross-section area $S$, made of a material of elasticity modulus $E$ and mass $\mu$ per length unit. The truss is clamped at the extremity $x = 0$ and free at the extremity $x = L$. We approximate the displacement field by a polynomial function of degree two. Taking into account the support condition, it reads:
\[ u (x) = \frac{x}{L}\, q_1 +  \left(\frac{x}{L} \right)^2 \, q_2
\]
where $q_1$ and $q_2$ are the components of the vector $\bm{q}$. The stiffness and mass matrix are:
\[ \bm{K} = \frac{E\, S}{L}\, \left[ {{\begin{array}{*{20}c}
		 1 & 1 \hfill \\
		1 & \frac{4}{3}\hfill \\
		\end{array} }} \right] , \qquad
	\bm{M} = \mu \, L \, \left[ {{\begin{array}{*{20}c}
		 \frac{1}{3} & \frac{1}{4} \hfill \\
		\frac{1}{4} & \frac{1}{5}\hfill \\
		\end{array} }} \right] 
\]
For sake of easiness, the units are choosen in such way that $E \,  S / L = \mu \, L = 1$. Solving the eigenvalue problem (\ref{H^2_omega u_i = lambda_i u_i}), we obtain two negative eigenvalues with multiplicity $2$:
\begin{equation}
   \lambda_1 = - 32.18070  , \quad
   \lambda_2 = - 2.48596
\label{lambda_1 & lambda_2 truss} 
\end{equation} 
In terms of of circular frequency $\omega_i = \sqrt{- \lambda_i}  $ and period $T_i$, we have $  \omega_1 = 5.672, \;    \omega_2 = 1.556, \;  T_1 = 1.107, \;  T_2 = 3.985$. The corresponding twin eigenvectors are for $\lambda_1$:
\begin{equation}
 \bm{u}_1 =\left[ {{\begin{array}{*{20}c}
		 -6.4220 \hfill \\
		8.8665 \hfill \\
		0.0 \hfill \\
		0.0 \hfill \\
		\end{array} }} \right], \qquad
		 \bm{v}_1 = 
\left[ {{\begin{array}{*{20}c}
		 0.0 \hfill \\
		0.0 \hfill \\
		0.075962 \hfill \\
		0.16780 \hfill \\
		\end{array} }} \right]
\label{u_1 & v_1 for the truss} 
\end{equation}
and for $\lambda_2$:
\begin{equation}
 \bm{u}_2 =\left[ {{\begin{array}{*{20}c}
		 -19.586 \hfill \\
		8.8665 \hfill \\
		0.0 \hfill \\
		0.0 \hfill \\
		\end{array} }} \right], \qquad
		 \bm{v}_2 = 
\left[ {{\begin{array}{*{20}c}
		 0.0 \hfill \\
		0.0 \hfill \\
		- 0.57233 \hfill \\
		- 0.41453 \hfill \\
		\end{array} }} \right]
\label{u_2 & v_2 for the truss} 
\end{equation}
They are of the form (\ref{u_j = (q_j 0) & v_j (0 M q_j)}) for a standard Hamiltonian.  They form a canonical basis of $V$.

That being said, we examine three data samples:
\begin{itemize}
\item \textbf{Sample 1: equilibrated combination of the two modes} ($\alpha_1 = \alpha_2 = 1$) as data:
\[ \bm{q} (t) = \left[ {{\begin{array}{*{20}c}
	    -6.4220 \hfill \\
		8.8665 \hfill \\
		\end{array} }} \right]\, \sin (5.672 \, t)
		+ \left[ {{\begin{array}{*{20}c}
		 -19.586 \hfill \\
		8.8665 \hfill \\
		\end{array} }} \right]\, \sin (1.556 \, t)
\]
We divide  the time interval form $0$ to $T$ into $m$ subintervals of same timestep. The snapshots are provided at the middle point of each subinterval. Computing the matrices (\ref{covariant snapshot matrix}) and  (\ref{B = C G C^T}), next solving the eigenvalue problem (\ref{H^2_omega u_i = lambda_i u_i}), we obtain two eigenvalues $\lambda_i$ of multiplicity $2$. Their numerical values are given in Table 1 for some values of $T$  and $m$. The last row  gives the reference values  (\ref{lambda_1 & lambda_2 truss}). The two latter columns provide the relative error $\mid \lambda'_i - \lambda_i \mid / \mid \lambda_i \mid $ with respect to these reference values $\lambda_i$. We  observe the convergence when increasing $T$ and $m$. 
\begin{table}[ht]
\begin{center}
\begin{small}
\begin{tabular}{*{6}{c}}
\hline
                &                & value of        & value of       & error on          & error on \\
 $T$         & $m$        & $\lambda'_1$ & $\lambda'_2$ & $ \lambda'_1$ & $\lambda'_2$ \\
\hline
10            & 10            & -31.83290 & -2.21129 & $1.08 \, 10^{-2}$ & $0.11$ \\
10            & 40            & -32.23037 & -2.46422 & $1.54 \, 10^{-3}$ &  $8.74 \, 10^{-3}$ \\
20            & 100          & -32.18026 & -2.48581 & $1.35 \, 10^{-5}$ &  $5.90 \, 10^{-5}$ \\
$+\infty$ & $+\infty$ & -32.18070 & -2.48596 &  -                          &   -   \\
\hline
\end{tabular}
\end{small}
\caption{sample $\alpha_1 = \alpha_2 = 1$, eigenvalues of $\bm{H}^2_\omega$}
\end{center}
\end{table}
The corresponding eigenvectors are given within a factor. After normalization according to the condition (\ref{normalization condition}), we obtain the approximations $\bm{u}'_i, \bm{v}'_i$ of the twin vectors. The relative errors:
\[ \parallel \bm{u}'_i - \bm{u}_i \parallel / \parallel \bm{u}_i \parallel , \qquad 
     \parallel \bm{v}'_i - \bm{v}_i \parallel / \parallel \bm{v}_i \parallel
\]
 with respect to the expected values (\ref{u_1 & v_1 for the truss}) and (\ref{u_2 & v_2 for the truss}) are given in Table 2 for $T = 20$ and $m = 100$. 
\begin{table}[ht]
\begin{center}
\begin{small}
\begin{tabular}{*{5}{c}}
\hline
eigenvectors       & $\bm{u}_1$ & $\bm{v}_1$ & $\bm{u}_2$ & $\bm{v}_2$ \\
\hline
relative error  & 3.18 \% & 0.64 \% & 0.88 \% & 0.02 \%  \\
\hline
\end{tabular}
\end{small}
\caption{sample $\alpha_1 = \alpha_2 = 1$,  error on the eigenvectors of $\bm{H}^2_\omega$}
\end{center}
\end{table}
The corresponding approximation of the Hessian matrix of the Hamiltonian is:
\begin{equation}
\bm{H} '=  \left[ {{\begin{array}{*{20}c}
		 0.99872 & 0.99634 & -3.7182 \, 10^{-3} & 4.6590 \, 10^{-3} \\ 
		 0.99634 & 1.3301 & 3.6907 \, 10^{-3} & -4.7010 \, 10^{-3} \\ 		 
		 -3.7182 \, 10^{-3} & 3.6907 \, 10^{-3} & 47.809 & -59.767 \\ 
		4.659 \, 10^{-3} & - 4.7010 \, 10^{-3} & -59.767 & 79.699 &  \\ 		
		\end{array} }} \right] 
\label{H' sample 1} 
\end{equation} 
to be compared to the exact value:
\begin{equation}
\bm{H} =  \left[ {{\begin{array}{*{20}c}
		 1.0 & 1.0 & 0.0 & 0.0 \\ 
		 1.0 & 1.3333 & 0.0 & 0.0 \\
		 0.0 & 0.0 & 48.0 & -60.0 \\
		0.0 & 0.0 & -60.0 & 80.0 \\
		\end{array} }} \right] 
\label{H truss} 
\end{equation} 
In the following matrix, the element at the intersection of the $\alpha$-th row and the $\beta$-th column is the relative error $\mid H'_{\alpha\beta} - H_{\alpha\beta}  \mid / \mid H_{\alpha\beta} \mid$  when it makes sense:
\[   \left[ {{\begin{array}{*{20}c}
		 0.127 \% & 0.365 \% & - & - \\ 
		 0.365 \% & 0.235 \% & - & - \\
		 - & - & 0.397 \% & 0.387 \% \\
		- & - & 0.387 \% & 0.375 \% \\
		\end{array} }} \right] 
\]
In a nutshell, when the data is the sum of the eigenmodes, we are able to deduce from the snapshot matrix a very accurate expression of the Hamiltonian.
\item \textbf{Sample 2: non equilibrated combination of the two modes} ($\alpha_1 = 1/2$, $\alpha_2 = 1$) as data:
\[ \bm{q} (t) = 0.5\,\left[ {{\begin{array}{*{20}c}
	    -6.4220 \hfill \\
		8.8665 \hfill \\
		\end{array} }} \right]\, \sin (5.672 \, t)
		+ \left[ {{\begin{array}{*{20}c}
		 -19.586 \hfill \\
		8.8665 \hfill \\
		\end{array} }} \right]\, \sin (1.556 \, t)
\]
In Table 3, we compare the numerical values of the eigenvalues $\lambda'_i$ to the reference values  $\lambda_i$  given by (\ref{g_i = alpha^2_i & k_i = & lambda_i =}).
\begin{table}[ht]
\begin{center}
\begin{small}
\begin{tabular}{*{6}{c}}
\hline
                &                & value of        & value of       & error on          & error on \\
 $T$         & $m$        & $\lambda'_1$ & $\lambda'_2$ & $ \lambda'_1$ & $\lambda'_2$ \\
\hline
20            & 100          & -2.01087 & -2.48629 & $2.07 \, 10^{-4}$ &  $1.34 \, 10^{-4}$ \\
$+\infty$ & $+\infty$ & -2.01129 & -2.48596 &  -                          &   -   \\
\hline
\end{tabular}
\end{small}
\caption{sample $\alpha_1 = 1/2$, $\alpha_2 = 1$,   eigenvalues of $\bm{H}^2_\omega$}
\end{center}
\end{table}
The twin eigenvectors   corresponding to $ \lambda'_1$ (resp. $\lambda'_2$) are very close to the reference values (\ref{u_1 & v_1 for the truss}), (\ref{u_2 & v_2 for the truss})).  The corresponding approximation of the Hessian matrix of the Hamiltonian is:
\[ \bm{H} '=  \left[ {{\begin{array}{*{20}c}
		 0.86145 & 0.69126 & 3.3236 \, 10^{-3} & 3.8656 \, 10^{-3} \\ 
		 0.69126 & 0.65214 & -6.1248 \, 10^{-4} & -6.9096 \, 10^{-4} \\ 		 
		 3.3236 \, 10^{-3}  &  -6.1248 \, 10^{-4} & 16.997 & -17.224 \\ 
		 3.8656 \, 10^{-3} & -6.9096 \, 10^{-4} & -17.224 & 20.957 &  \\ 		
		\end{array} }} \right] 
\]
By comparison to the exact value (\ref{H truss}), we observe that the matrix is deteriorated, although its global structure is conserved. 
\item \textbf{Sample 3: only the second eigenmode} ($\alpha_1 = 0, \alpha_2 = 1$) as data:
\[ \bm{q} (t) = \left[ {{\begin{array}{*{20}c}
		 -19.586 \hfill \\
		8.8665 \hfill \\
		\end{array} }} \right]\, \sin (1.556 \, t)
\]
With $T = 20$ and $m = 100$, we obtain two eigenvalues of multiplicity $2$:
\[ \lambda'_1 = -5.5511 \cdot 10^{-17} , \qquad
     \lambda'_2 = -2.48592
\]
In terms of absolute values, the latter eigenvalue overwhelms the former one, that is expected because the data are provided only by the second eigenmode. 
The corresponding approximation of the Hessian matrix of the Hamiltonian is:
\[ \bm{H} '=  \left[ {{\begin{array}{*{20}c}
		 0.81736 & 0.59201 & -2.0777 \, 10^{-4} & 9.4055 \, 10^{-5} \\ 
		0.59201 & 0.42879 & -1.5048 \, 10^{-4} & 6.8124\, 10^{-5} \\ 		 
		-2.0777 \, 10^{-4}   & -1.5048 \, 10^{-4} & 6.7325 & -3.0477 \\ 
		9.4055 \, 10^{-5}  & 6.8124\, 10^{-5} & -3.0477 & 1.3796 &  \\ 		
		\end{array} }} \right] 
\]
to be compared to the exact value (\ref{H truss}). As predictable, the deterioration is larger than for the value  (\ref{H' sample 1}) given by the sample 2 but not so much. Besides, the twin eigenvectors corresponding to $ \lambda'_1$ (resp. $\lambda'_2$) are very close to the reference values. 

\section{Conclusions and perspectives}

The main interest of the method is, for large scale systems, to extract from experimental data or numerical simulations the principal component of the Hamiltonian, operation that can be done offline from a big data cloud. Next, the reduced system can be used online to predict the response to given excitations by solving a canonical equation system of small size. Besides, whenever the data cloud is enriched, the Hamiltonian can be updated, according to the machine learning process. 

The application realm of the proposed method is not limited to the structural mechanics but can be extended to the homogenization of materials to find the effective properties by considering a reference elementary volume \cite{Willis 2012}.
In the future, we hope   to extend the approach also to dissipative dynamical systems, first in the linear case of damping, next to the nonlinear case of elastoplasticity and viscoelastoplasticity. 

\end{itemize}
 

%
%
%

\begin{thebibliography}{8}

\bibitem{Chinesta 2014}
Chinesta, F., Ladev\`{e}ze, P.: Separated Representations and PGD-Based Model Reduction, Fundamentals and Applications. Springer (2014)


\bibitem{Geradin 2015}
G\'eradin, M., Rixen, D.: Mechanical Vibrations: Theory and Application to Structural Dynamics. 3rd edn. Wiley (2015)

\bibitem{Libermann_1987}
Libermann, P., Marle, C.-M.: Symplectic Geometry and Analytical 
Mechanics. D. Reidel Publishing Company, Dordrecht (1987)

\bibitem{Peng 2016}
Peng, L., Mohseni, K.: Symplectic model reduction of Hamiltonian systems, SIAM Journal of Scientific Computing, \textbf{38}(1), A1--A27 (2016)

\bibitem{Souriau_1997b} 
Souriau, J.-M.: Structure of Dynamical Systems, a Symplectic View of Physics. Birkh\"{a}user Verlag, New York, (1997)

\bibitem{Willis 2012}
Willis, J.: The construction of effective relations for waves in a composite, C.R. Mecanique, \textbf{340}, 181-192 (2012)

\end{thebibliography}
%

\end{document}